\documentclass[superscriptaddress,aps,nofootinbib,preprint]{revtex4-1}
\usepackage{graphicx}
\usepackage{color}
\usepackage{amsmath,amssymb}  % Better maths support & more symbols
\usepackage{bm}  % Define \bm{} to use bold math fonts
\usepackage{bbm}  % Define \bm{} to use bold math fonts
\usepackage{slashed}

\newcommand\beq{\begin{equation}}
\newcommand\eeq{\end{equation}}
\newcommand\Tr{\text{Tr}\,}
\newcommand\nn{\nonumber}

\newcommand\hc{\text{h.c.}}
\newcommand\PD{{\phantom{\dagger}}}
\newcommand\PP{{\phantom{\prime}}}
\newcommand\LLH{$\text{L}^2\text{H}$}
\newcommand\sfrac[2]{{\textstyle{\frac{#1}{#2}}}}

\def\OMIT#1{{}}

\begin{document}

%%%%%%%%%%%%%%%%%Title%%%%%%%%%%%%%%%%%%%%%%%
\title{One Loop Renormalization of the Littlest Higgs Model}

\author{Benjam\'\i{}n Grinstein}
\email[]{bgrinstein@ucsd.edu}
\affiliation{Department of Physics, University of California at San Diego, La Jolla, CA 92093}
\affiliation{Theory Division, CERN, CH-1211 Geneva 23, Switzerland}

\author{Randall Kelley}
\email[]{randallkelley@physics.harvard.edu}
\affiliation{Center for the Fundamental Laws of Nature, Harvard University, Cambridge, MA 02138}

\author{Patipan Uttayarat}
\email[]{puttayarat@physics.ucsd.edu}
\affiliation{Department of Physics, University of California at San Diego, La Jolla, CA 92093}

\preprint{UCSD PTH 10-11}

%%%%%%%%%%%%%%%%Abstract%%%%%%%%%%%%%%%%%%%%%%%
\begin{abstract}
  In Little Higgs models a collective symmetry prevents the Higgs from
  acquiring a quadratically divergent mass at one loop. This
  collective symmetry is broken by weakly gauged interactions. Terms,
  like Yukawa couplings, that display collective symmetry in the bare
  Lagrangian are generically renormalized into a sum of terms
  that do not respect the collective symmetry except possibly at one
  renormalization point where the couplings are related so that the
  symmetry is restored. We study here the one loop renormalization of a
  prototypical example, the Littlest Higgs Model. Some features of the
  renormalization of this model are novel, unfamiliar form similar
  chiral Lagrangian studies. 
\end{abstract}

\maketitle

%----------------------------------------------------------------------
\section{Introduction}
The Littlest Higgs (\LLH) model \cite{ArkaniHamed:2002qy} is a
realization of the idea that the Higgs field, responsible for
electroweak symmetry breaking, is a pseudo-Goldstone boson, and as
such its mass is automatically small (for some reviews see
Ref.~\cite{Schmaltz:2005ky}). What is meant by ``small'' is that the
Higgs mass can be made arbitrarily small compared to the scale of
breaking of the symmetry that gives rise to this Goldstone boson.
Earlier realizations of this idea faced difficulties, required
additional fine tuning~\cite{Kaplan:1983fs}. 
In the \LLH\ model, as well as its many extensions, the absence of
quadratically divergent radiative corrections to the Higgs mass is
guaranteed, at one loop order, by the collective symmetry
argument. The argument fails beyond one loop order, so the Higgs can
be made naturally light only if its mass is no smaller than of the
order of a two loop radiative correction with a cut-off at the scale of the
new physics.

While there is a vast literature exploring the phenomenological
effects of \LLH-type models, the renormalization structure
of the model has been little explored. Computations have been
presented that check that the collective symmetry argument does
work; however, the structure of counterterms needed to subtract the 
divergences that do occur has not been studied. 
Furthermore, the renormalization group equations have not been determined. 

Phenomenologically the \LLH\ model has fallen somewhat out of favor
because of its difficulties simultaneously accommodating the
electroweak precision constraints and in 
solving the little hierarchy problem
\cite{Han:2003wu}. However, its structure is prototypical of many models,
like Littlest Higgs models with reduced gauge symmetry
\cite{Perelstein:2003wd}, or with custodial \cite{Chang:2003un} or
T-parity \cite{Cheng:2003ju} symmetries. Therefore, the methods we
will introduce here should be directly applicable to the one loop 
renormalization of any of the models in this class.

It was noted in Ref.~\cite{us} that renormalization group running of the top
Yukawa coupling in \LLH-models disrupts the collective
symmetry. That is,  in order for the collective symmetry argument to operate in
the top-quark Yukawa sector, the coupling is built to satisfy an
$SU(3)$ symmetry. However, this symmetry is broken by weak gauge
interactions. The would be $SU(3)$ symmetric top-Yukawa coupling
actually splits into two $SU(2)\times U(1)$ symmetric terms with
coupling constants that run away from each other as they evolve under
the renormalization group. This begs the question, what is the full
renormalization group structure of the model? It is the purpose of
this paper to address this question, at one loop order. 

There are several energy scales associated with this model. In
addition to the cutoff, $\Lambda$, there is the scale of masses of heavy vector
bosons, $gf$ where $f\sim \Lambda/4\pi$ is a Goldstone boson decay
constant and $g$ some gauge coupling, and the electroweak
breaking scale $v$.  We are largely
interested in the cut-off dependence, so for our computations we will
focus on the largest energies, above $gf$. Therefore to determine the ultraviolet
behavior we retain the massive gauge vector bosons in our calculations
and neglect their masses. On the other hand, the renormalization
structure below the scale of these masses, $gf$, is well
understood. The model reduces there to the standard electroweak model
with one Higgs doublet supplemented by irrelevant operators. 

The main result of this paper, the splitting of the Yukawa couplings
responsible for the top quark mass, was already noted in
Ref.~\cite{us}. There, a no-go theorem for the collective symmetry
mechanism for Yukawa terms was proved. However,  the details of the
calculation of the running of Yukawa couplings were not given there
since, as can be seen from this work, this merits a lengthy discussion
that would have detracted from its main point.  In fact, we have
encountered several stumbling blocks, and corresponding solutions,
along the way.  Readers interested in questions of principle or
practice, or both, in \LLH-type models, will hopefully find
this work useful.

The paper is organized as follows. We first review the \LLH\ Model
in Sec.~\ref{sec:model}. We classify the counterterms needed to
renormalize the model at one loop in Sec.~\ref{sec:counterterms} and
proceed to compute the renormalization constants and corresponding beta
functions in Sec.~\ref{sec:renormalization}. We offer some brief
concluding remarks in Sec.~\ref{sec:conc}.

\section{The Model}
\label{sec:model}
The \LLH\ model is an effective low energy description
of some incompletely specified shorter distance dynamics. The short
distance dynamics has a global ``flavor'' symmetry $G_f=SU(5)$, of
which a subgroup $G_w=SU(2)\times SU(2)\times U(1)\times U(1)$ is
weakly gauged. In the absence of this weak gauge force, the flavor
symmetry is broken spontaneously to a subgroup $H=SO(5)$ due to
hyper-strong interactions at a scale $\Lambda$. As a result, there are
massless Goldstone bosons that are coordinates on the $G_f/H$ coset
space. Since the weakly gauged $G_w$ force breaks the flavor symmetry
explicitly, including its effects leads to some of the Goldstone
bosons (the would-be Goldstone bosons) being eaten by the Higgs
mechanism and the rest becoming pseudo-Goldstone bosons (PGBs)
acquiring small masses of order $\Lambda$ times a small symmetry
breaking parameter, a gauge coupling constant of the weakly gauged
$G_w$.  The Higgs is the lightest PGB in Little Higgs models, and its mass is
naturally much less than $\Lambda$ (and the other PGBs): due to the
collective symmetry breaking mechanism a contribution of order
$\Lambda^2$ to its mass arises only at two loops.

To establish notation we briefly review elements of the \LLH.  Symmetry breaking $SU(5)\to
SO(5)$ is characterized by the Goldstone boson decay constant $f$. The
embedding of $G_w$ in $G_f$ is fixed by taking the gauge generators 
\begin{equation}
\label{eq:gauge-generator}
\begin{split}
	Q_1^a &= \begin{pmatrix} \tau^a/2 &0 &0\\ 0 &0 &0\\ 0 &0 &0\end{pmatrix},\quad Y_1 = \text{diag}(3,3,-2,-2,-2)/10,\\
	Q_2^a &= \begin{pmatrix} 0 &0 &0\\ 0 &0 &0\\ 0 &0 &-\tau^{a*}/2\end{pmatrix},\quad Y_2 = \text{diag}(2,2,2,-3,-3)/10.
\end{split}
\end{equation}

The vacuum manifold is characterized by a unitary, symmetric
$5\times5$ matrix $\Sigma$, transforming as $\Sigma\to U\Sigma U^T$
under $U\in SU(5)$. A convenient parametrization of $\Sigma$ in terms
of the hermitian matrix of Goldstone bosons $\Pi$ is 
\begin{gather}
 \label{eq:Sigma}
\Sigma=e^{2i\Pi/f}\Sigma_0,\quad  
  \Sigma_{0}=\begin{pmatrix} 0&0& \mathbbm{1}_{2\times2}\\ 0 & 1& 0\\
    \mathbbm{1}_{2\times2} &0 &0\end{pmatrix},\\
\intertext{where}
\Pi=\begin{pmatrix} \omega+\eta\mathbbm{1}/\sqrt{20}&h/\sqrt2& \phi\\ h^\dagger/\sqrt2 & -2\eta/\sqrt{5}& h^T/\sqrt2\\
   \phi^* &h^*/\sqrt2 &\omega^T+\eta\mathbbm{1}/\sqrt{20}\end{pmatrix}
\end{gather}
Here $\Sigma_0$ gives the dynamically determined direction in which
the vacuum aligns\footnote{To ensure this alignment the weakly gauge
  coupling constant have to be strong enough; see
  Ref.~\cite{Grinstein:2008kt}.} \cite{Preskill:1980mz} relative to the
embedding of $G_w$ in $G_f$ given in
Eq.~\eqref{eq:gauge-generator}. Fluctuations along broken symmetry directions
are parametrized by fourteen fields in $\Pi$: $\omega$ and $\phi$ are
$2\times2$ matrices satisfying $\omega^\dagger=\omega$ and
$\phi^T=\phi$, $h$ is an unrestricted $2\times1$ matrix and $\eta$ is
$1\times1$ and real. The vacuum spontaneously breaks $G_w\to
SU(2)\times U(1)$, and the four fields in $\omega$ and $\eta$ are
eaten by the broken  generators of gauge symmetries.

The covariant derivative is
\begin{equation}
	D_\mu\Sigma = \partial_\mu\Sigma - i\sum_{j=1}^2[g_jW^a_{j\mu}(Q^a_j\Sigma+\Sigma Q_j^{aT})+g^\prime_jB_{j\mu}(Y_j\Sigma+\Sigma Y_j)],
\end{equation} 
where $B_j$ and $W^a_j$ are the $U(1)_j$ and $SU(2)_j$ gauge fields
respectively.  The $U(1)_j$ coupling constant is taken to be $g_j^\prime$
while the $SU(2)_j$ coupling constant is $g_j$.

The effective low energy theory has kinetic term 
\begin{equation}
	\mathcal{L}_{kin} =  \frac{f^2}{8}\text{Tr}(D_\mu\Sigma)(D^\mu\Sigma)^\dagger. \label{eq:Lkin}
\end{equation}
If one sets $g_{1}=g_1'=0$ the model has an exact global
$SU(3)$ symmetry (acting on upper $3 \times 3$ block of $\Sigma$),
while for $g_{2}=g_2'=0$ it has a different exact global $SU(3)$
symmetry (acting on the lower $3 \times 3$ block).  Either of these
exact global $SU(3)$ would-be symmetries guarantee the Higgs remains
exactly massless. Hence, the Higgs mass should vanish for either
$g_{1}=g_1'=0$ or $g_{2}=g_2'=0$. The perturbative quadratically
divergent correction to the Higgs mass must be polynomial in the
couplings and can involve only one of the couplings at a time at one loop
order. Hence it must vanish at one loop. This is the collective
symmetry mechanism that ensures the absence of 1-loop quadratic
divergences in the Higgs mass.

For a top-quark sector introduce a pair of singlet Weyl fermions $u_L$
and $u_R$ with hypercharge 2/3.  $u_L$ is combined with the
$3^\text{rd}$ generation doublet $q_L=(t_L, b_L)^T$ to form a
``royal'' triplet
\begin{equation}
	\chi_L = \begin{pmatrix}i\tau^2 q_L\\u_L\end{pmatrix}.
\end{equation}
The top Yukawa interaction is obtained from coupling the fermions to
the upper right $2 \times 3$ block of the $\Sigma$ field,
\begin{equation}
  \label{LTop}	
  \mathcal{L}_{top} = -\frac12\lambda_1f\bar \chi_{L
    I}\epsilon^{IJK}\epsilon^{xy}\Sigma_{Jx}\Sigma_{Ky}q_R 
  - \lambda_2f\bar u_L u_R + h.c.
\end{equation}
Here and  below implicit sums are over 1, 2, 3 for $I,J,K$, over 1,
2, for $i,j,k$ and over 4, 5 for $x,y$.

There is in fact no symmetry reason for the fields in $\chi_L$ to
combine into a triplet \cite{us}. More generally the coupling is of the
form
\begin{equation}
\label{eq:Ltop-split}
{\cal L}_{\text{top}}=-\lambda_1f\bar \chi_{Li}\epsilon^{ij}\epsilon^{xy}
\Sigma_{jx}\Sigma_{3y}q_R
-\frac12\lambda'_1f\bar u_{L}\epsilon^{jk}\epsilon^{xy}
\Sigma_{jx}\Sigma_{ky}q_R
-\lambda_2f\bar u_Lu_R +\hc
\end{equation}
In this case, there is a quadratically divergent correction to the Higgs mass, 
\begin{equation}
\label{oops}
\delta m_h^2
=\frac{6}{16\pi^2}(\lambda^2_1-\lambda^{\prime2}_1)\Lambda^2
\end{equation}
where $\Lambda$ is a UV cut-off. As we will show below the relation
$\lambda_1'=\lambda_1$ is unstable against radiative
corrections.

%----------------------------------------------------------------------

\section{General Structure of  Counterterms}
\label{sec:counterterms}
\subsection{Scalar Kinetic Energy Counterterms}
Kinetic energy counterterms are normally introduced in field theory by
rescaling the bare fields $\phi\to Z^{1/2}\phi$. In non-linear sigma
models the self-interactions of Goldstone bosons require counterterms
that are higher order in the derivative expansion, and no rescaling of
fields is necessary. However, non-linear sigma models coupled to light
gauge bosons and fermions do generally require counterterms quadratic
in derivatives. We will see  that in the \LLH\ model no rescaling
$\phi\to Z^{1/2}\phi$ is needed. Instead new terms that are not
symmetric under the full $SU(5)$ symmetry are required to completely
subtract the model at one loop.

We begin our study of the structure of kinetic energy counterterms by
considering the slightly simpler case $\lambda_1'=\lambda_1$.  Working
only to 1 loop, there is only one coupling constant present in each
divergent self-energy diagram so the corresponding counterterm could
just as well be computed setting all other coupling constants to zero.
The Lagrangian with all but one couplings set to zero has an
$SU(3)\times SU(2)\times U(1)$
symmetry. Since we can choose the regulator to respect this symmetry
we demand the counterterms are invariant under $SU(3)\times
SU(2)\times U(1)$. 

Consider  the possibility of partly subtracting the divergent graphs
by rescaling
the bare fields. In general we can choose a different wavefunction
renormalization factor $Z$ for each of the fourteen Goldstone boson
fields in $\Pi$. Were the interaction and the regularization method to
respect the full flavor symmetry ($SU(5)$), there would only be one
common $Z$ for all the fields in $\Pi$. The question  becomes:
what is the restriction that $SU(3)\times SU(2)\times U(1)$ imposes on the $Z$?

To answer this consider the expansion of the bare kinetic term
\begin{equation}
\frac{f^2}{8}\Tr \partial_\mu\Sigma^{\dagger}\partial^\mu\Sigma
= \Tr \partial_\mu\phi^{\dagger}\partial^\mu\phi
+\sfrac12 \partial_\mu\eta\partial^\mu\eta
+\Tr \partial_\mu\omega\partial^\mu\omega
+\partial_\mu h^{\dagger}\partial^\mu h +\dots
\end{equation}
where the ellipsis stand for terms quartic in the fields. Now we
rescale each of the fourteen fields by an independent factor $Z$ and
ask what are the constraints from imposing $SU(3)\times SU(2)\times
U(1)$.  There is a $SU(2)\times U(1)$ subgroup that acts linearly and
hence there are only four different $Z$ factors:
\begin{equation}
\label{eq:kin-term}
Z_\phi\Tr \partial_\mu\phi^{\dagger}\partial^\mu\phi
+\sfrac12 Z_\eta\partial_\mu\eta\partial^\mu\eta
+Z_\omega\Tr \partial_\mu\omega\partial^\mu\omega
+Z_h\partial_\mu h^{\dagger}\partial^\mu h +\dots
\end{equation}
We are led to consider the restrictions from $SU(3)$ on these four
factors. It is a straightforward but laborious exercise to compute the
transformation properties of the fields in $\Pi$ under $SU(3)$. We
take for definiteness the $SU(3)$ generated by the top-left $3\times3$
block. Of particular interest are transformations generated by the 4-7
Gell-Mann matrices
\begin{equation}
\sum_{a=4}^7\epsilon^a
T^a=\sum_{a=4}^7\epsilon^a \begin{pmatrix}\lambda^a&0_{3\times2}\\0_{2\times3}&0_{2\times2}\end{pmatrix}
\equiv \begin{pmatrix}0_{2\times2}&\lambda&0_{2\times2}\\
\lambda^\dagger&0&0_{1\times2}\\
0_{2\times2}&0_{2\times1}&0_{2\times2}\end{pmatrix},
\end{equation}
where $\lambda$ is a $2\times1$ complex matrix of order $\epsilon$. The resulting nonlinear transformations, to first order in $\epsilon$, are 
\begin{align}
\delta h &= \frac1{\sqrt2}f\lambda
+\frac{i}{\sqrt2}\left[-\omega\lambda-\frac5{\sqrt{20}}\eta\lambda+\phi\lambda^*\right]+\cdots\\
\delta\phi &=\frac{i}{2\sqrt2}\left[h\lambda^T+\lambda h^T\right]+\cdots\\
\delta\eta
&=i\frac{\sqrt{10}}{4}\left[h^\dagger\lambda-\lambda^\dagger
  h\right]+\cdots\\
\delta\omega&=\frac{i}{2\sqrt2}\left[\lambda h^\dagger -h\lambda^\dagger\right]-
\frac{i}{4\sqrt2}\left[h^\dagger\lambda-\lambda^\dagger
  h\right]\mathbf{1}+\cdots
\end{align}
where the ellipses stand for terms of quadratic and higher  order in
the fields.

Applying this variation to the kinetic term in \eqref{eq:kin-term} and
retaining only terms quadratic in the fields we obtain
\begin{multline}
  \delta{\cal L}=\frac1{\sqrt2}(Z_\phi-Z_h)\Tr \partial_\mu\phi^{\dagger}\partial^\mu h\lambda^T +\hc \nonumber\\
+\frac1{\sqrt2}(Z_\omega-Z_h)\Tr \partial_\mu\omega\partial^\mu\left[\lambda  h^\dagger -h\lambda^\dagger\right]+\frac{\sqrt{10}}{4}(Z_\eta-Z_h)\Tr \partial_\mu\eta\partial^\mu\left[h^\dagger\lambda-\lambda^\dagger  h\right] \nonumber
\end{multline}
Hence invariance under $SU(3)$ requires
$Z_h=Z_\phi=Z_\omega=Z_\eta\equiv Z$. The same conclusion is
reached by consideration of other embeddings of the invariance
subgroup.

Already in the special $SU(3)\times SU(2)\times U(1)$-symmetric  case
one sees that  divergences in the self-energy diagrams cannot be subtracted
with a single common $Z$ factor. One must  introduce
counterterms invariant under $SU(3)\times SU(2)\times U(1)$, or more
generally, under $G_w$, that are  not
invariant under $SU(5)$. We next turn to constructing the relevant counterterms.

%----------------------------------------------------------------------

\subsubsection{Scalar Kinetic Counterterms from Gauge Interaction}

\begin{figure}[htbp]
\begin{center}
\includegraphics[width=0.5\textwidth]{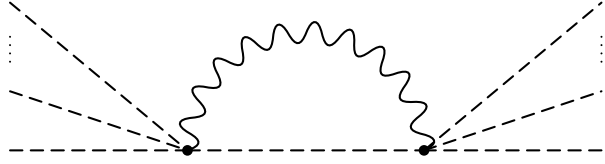}
\caption{Scalar 2-point Function from gauge interaction with background pion fields.}
\label{fig:scalar2pt-gauge}
\end{center}
\end{figure}

Gauge interactions induce divergences in the scalar 2-point function
with arbitrary background pion fields as shown in
Fig.~\ref{fig:scalar2pt-gauge}. We obtain the counterterms by the
method of spurions.  The gauge generators are promoted to
spurions transforming in the adjoint representation of $SU(5)$,
$T^a\to UT^aU^\dagger$. We list all the $SU(5)$ invariant
counterterms with two $T^a$'s and two derivatives.   In the
$SU(2)_1$ sector we find
\begin{equation}
\begin{aligned}
  \mathcal{O}^{g_{1}}_{1} &= \Tr\left(Q_{1}^aQ_{1}^a\right)\Tr\left(\partial_\mu\Sigma\partial^\mu\Sigma^*\right),\\
  \mathcal{O}^{g_{1}}_{2} &= \Tr\left(Q_{1}^aQ_{1}^a\partial_\mu\Sigma\partial^\mu\Sigma^*\right),\\
  \mathcal{O}^{g_{1}}_{3} &= \Tr\left(Q_{1}^a\partial_\mu\Sigma(Q_{1}^a)^T\partial^\mu\Sigma^*\right),\\
  \mathcal{O}^{g_{1}}_4 &=\Tr\left(Q_{1}^a(\partial_\mu\Sigma)\Sigma^*\right)\Tr\left(Q_{1}^a\Sigma\partial^\mu\Sigma^*\right),\\
  \mathcal{O}^{g_{1}}_5 &=\Tr\left(Q_{1}^a(\partial_\mu\Sigma)\Sigma^*Q_{1}^a\Sigma\partial^\mu\Sigma^*\right),\\
  \mathcal{O}^{g_{1}}_6 &=\Tr\left(Q_{1}^a\Sigma(Q_{1}^a)^T\Sigma^*\right)\Tr\left(\partial_\mu\Sigma\partial^\mu\Sigma^*\right),\\
  \mathcal{O}^{g_{1}}_7
  &=\Tr\left(Q_{1}^a(\partial_\mu\Sigma)(\partial^\mu\Sigma^*)\Sigma(Q_{1}^a)^T\Sigma^*\right)
  + \hc.
\end{aligned}
\end{equation}
The counterterms for the $SU(2)_2$ sector are obtained from those in
the $SU(2)_1$ sector by the replacements $g_1\to g_2$ and $Q^a_1\to
Q^a_2$. For the $U(1)_1$ sector we have
\begin{equation}
\begin{aligned}
	\mathcal{O}^{g'_{1}}_1 &= \Tr\left(Y_{1}Y_{1}\right)\Tr\left(\partial_\mu\Sigma\partial^\mu\Sigma^*\right)\\
	\mathcal{O}^{g'_{1}}_2 &= \Tr\left(Y_{1}Y_{1}\partial_\mu\Sigma\partial^\mu\Sigma^*\right)\\
	\mathcal{O}^{g'_{1}}_3 &= \Tr\left(Y_{1}\partial_\mu\Sigma Y_{1}\partial^\mu\Sigma^*\right)\\
	\mathcal{O}^{g'_{1}}_4 &=\Tr\left(Y_{1}(\partial_\mu\Sigma)\Sigma^*\right)\Tr\left(Y_{1}\Sigma\partial^\mu\Sigma^*\right)\\
	\mathcal{O}^{g'_{1}}_5 &=\Tr\left(Y_{1}(\partial_\mu\Sigma)\Sigma^*Y_{1}\Sigma\partial^\mu\Sigma^*\right)\\
	\mathcal{O}^{g'_{1}}_6 &=\Tr\left(Y_{1}\Sigma Y_{1}\Sigma^*\right)\Tr\left(\partial_\mu\Sigma\partial^\mu\Sigma^*\right)\\
	\mathcal{O}^{g'_{1}}_7 &=\Tr\left(Y_{1}(\partial_\mu\Sigma)(\partial^\mu\Sigma^*)\Sigma Y_{1}\Sigma^*\right) + \hc.
\end{aligned}
\end{equation}
Similarly, the counterterms for the $U(1)_2$ sector can be obtained by
substituting $g^\prime_1\to g^\prime_2$ and $Y_1\to Y_2$ in the
operators above. 
%----------------------------------------------------------------------

\subsubsection{Scalar Kinetic Counterterms from Yukawa Interaction}

\begin{figure}[htbp]
\begin{center}
\includegraphics[width=0.5\textwidth]{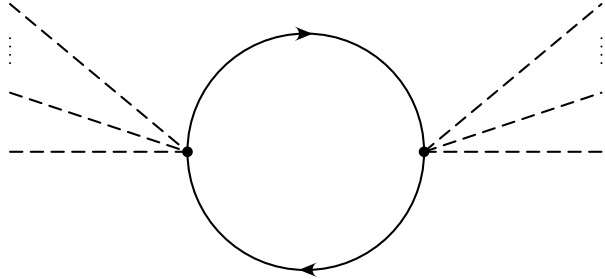}
\caption{Scalar 2-point Function from Yukawa interaction with background pion fields.}
\label{fig:scalar2pt-yukawa}
\end{center}
\end{figure}

Yukawa interactions also induce divergences in the scalar 2-point
function with arbitrary background pion fields as shown in
Fig.~\ref{fig:scalar2pt-yukawa}. Just as was done for gauge
generators, we treat the Yukawa couplings as 
$SU(5)$ breaking spurions.  In doing so, we promote $\chi_L$ to a
5-plet
\begin{equation}
  \mathcal{L}_{Yuk} = \bar \chi_{La}^\PD S^{abcde}\Sigma_{bc}\Sigma_{de} q_R + \hc,
\end{equation}
with $S$ symmetric in $\{b,c\}$, $\{d,e\}$ and the exchange of the
pair $(b,c) \leftrightarrow (d,e)$.  The spurion $S$ is not arbitrary,
but rather  takes a fixed ``vacuum expectation'' value
\begin{equation}
  <S^{abcde}> =
  \left\{ \begin{array}{ll}\frac{\lambda_1}{8}\epsilon^{abd45}\epsilon^{123ce}+\ldots
      \quad a=1,2\\ 
      \frac{\lambda_1^\prime}{8} \epsilon^{3bd45}\epsilon^{123ce}+\ldots\quad a=3\end{array}\right.
\end{equation}
where $+\ldots$ stands for symmetrization.  Note that we can demand
that $S \rightarrow S^\ast$ under CP, so $\mathcal{L}$ is invariant
under CP.  The counterterms will be also invariant under CP and hence
hermitian. For notational compactness we define $\Psi^{a} =
S^{abcde}\Sigma_{bc}\Sigma_{de}$.  In terms of this, the counterterm is
\begin{equation}
	\label{eq:yukawa-ct}
	\mathcal{O}_\Psi
		= \partial_\mu \Psi^{\dagger a}\partial^\mu \Psi_a.
\end{equation}

%----------------------------------------------------------------------

\begin{figure}[htbp]
\begin{center}
\includegraphics[width=0.5\textwidth]{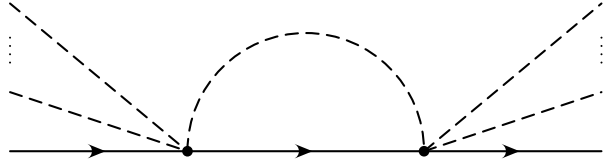}
\caption{Fermion 2pt Function from Yukawa interaction with background pion fields}
\label{fig:fermion2pt}
\end{center}
\end{figure}

\subsection{Fermion Kinetic Energy Counterterms}

The divergence in the fermion self-energy is also present in the
diagram with arbitrary number of pion fields at each of the Yukawa
vertices, as shown in Fig.~\ref{fig:fermion2pt}.  For notational
compactness we defined $\Psi^{abc} = S^{abcde}\Sigma_{de}$ and
$\xi_{abc} = S^\ast_{abcde}\Sigma^{\ast de}$.  The counterterms for
the $q_R$ 2-point function are
\begin{equation}
\begin{aligned}
  \mathcal{O}_{q_1} &= \bar q_R\bar{\Psi}_{abc} i \slashed{\partial}\Psi^{abc}q_R,\\
  \mathcal{O}_{q_2} &= \bar q_R\bar{\Psi}_{abc} \Sigma^{\ast ce} i \slashed{\partial} \Sigma_{ed} \Psi^{abd}q_R,\\
  \mathcal{O}_{q_3} &= \bar q_R\bar{\Psi}_{abc} \Sigma^{\ast bc}i \slashed{\partial} \Sigma_{de}\Psi^{ade}q_R,\\
  \mathcal{O}_{q_4} &= \bar q_R\bar{\Psi}_{abc} \Sigma_{de}i \slashed{\partial} \Sigma^{\ast bc}\Psi^{ade}q_R,\\
  \mathcal{O}_{q_5} &= \bar q_R\bar{\Psi}_{abc} \gamma^\mu
  \Psi^{ade} \partial_\mu \left(\Sigma^{\ast bc}\Sigma_{de}\right)q_R,
\end{aligned}
\end{equation}
while the counterterms for $\chi_L$ 2-ponit function are
\begin{equation}
\begin{aligned}
  \mathcal{O}_{\chi_1} &= \bar\chi_{La}\bar\xi^{abc} i \slashed{\partial}\xi_{a^\prime bc}\chi_L^{a^\prime},\\
  \mathcal{O}_{\chi_2} &= \bar\chi_{La}\bar\xi^{abc} \Sigma_{ce} i \slashed{\partial} \Sigma^{\ast ed} \xi_{a^\prime bd}\chi_L^{a^\prime},\\
  \mathcal{O}_{\chi_3} &= \bar\chi_{La}\bar\xi^{abc} \Sigma_{ bc}i \slashed{\partial} \Sigma^{\ast de}\xi_{a^\prime de}\chi_L^{a^\prime},\\
  \mathcal{O}_{\chi_4} &= \bar\chi_{La}\bar\xi^{abc} \Sigma^{\ast de}i \slashed{\partial} \Sigma_{bc}\xi_{a^\prime de}\chi_L^{a^\prime},\\
  \mathcal{O}_{\chi_5} &= \bar\chi_{La}\bar\xi^{abc} \gamma^\mu
  \xi_{a^\prime de} \partial_\mu \left(\Sigma^{\ast
      de}\Sigma_{bc}\right)\chi_L^{a^\prime}.
\end{aligned}
\end{equation}
%----------------------------------------------------------------------

\subsection{Yukawa Vertex Counterterms}
At 1-loop order, gauge interactions do not introduce a new
counterterm.  So we can subtract off the divergences with the Yukawa
operator ({\it i.e.,} $\bar \chi_{La}^\PD
S^{abcde}\Sigma_{bc}\Sigma_{de} q_R + \hc$).  This is not the case for
Yukawa interactions which generate two new counterterms
\begin{equation}
  \begin{aligned}
    \mathcal{O}_{v_1} &= \bar q_RS^\ast_{abcde}\Sigma^{\ast de}S^{lmnop}\Sigma_{mn}\Sigma_{op}S^\ast_{lqrst}\Sigma^{\ast st} \Sigma^{\ast bc}\Sigma^{\ast qr}\chi_L^a+\hc,\\
    \mathcal{O}_{v_2} &= \bar q_RS^\ast_{abcde}\Sigma^{\ast
      de}S^{lmnop}\Sigma_{mn}\Sigma_{op}S^\ast_{lqrst}\Sigma^{\ast st}
    \Sigma^{\ast bq}\Sigma^{\ast cr}\chi_L^a+\hc
  \end{aligned}
\end{equation}

%----------------------------------------------------------------------

\subsection{Counterterms to counterterms: The general case}
The counterterms displayed so far are appropriate to render all green
functions finite if the only interactions in the model are those
displayed in the Lagrangian given in Sec.~\ref{sec:model}. That is,
the counterterms are appropriate to the case were the bare Lagrangian
has the form given in Sec.~\ref{sec:model}. However, this cannot be
maintained beyond 1-loop order. The 1-loop counterterms become
interaction terms at 2-loops. This requires additional 
counterterms. And so on, as one moves to higher orders in the loop
expansion. All terms consistent with the symmetries of the model will
be generated by renormalization. 

It is more natural to start with the complete set of interaction terms
(formerly counterterms) and treat them all on an equal footing. However, this
is not a viable program for this model since the complete set does not
appear to be finite. The next best option is an organizing principle
for a calculation that requires finite precision. 

Before we make a specific proposal for one such organizing principle,
we would like to contrast this with other models. Clearly
the case of renormalizable theories is very different: only a
finite number of terms is required to renormalize the theory to all
orders in the loop expansion. More apropos, the case of chiral
Lagrangians is different too. For these as one goes up in the loop
expansion the counterterms involve accordingly more
derivatives. Therefore the infinite set of counterterms are neatly
organized by the number of derivatives which is tied to the loop expansion. In
the \LLH\ this is explicitly not the case already at 1-loop order: the
counterterms generated are not suppressed by additional derivatives. 

Suppose we are interested in processes that do not involve more than
$n$ PGBs. By expanding the $\Sigma$ field in powers of the PGBs we
will discover there is a finite number, $N(n,d)$ of linearly
independent operators containing no more than $d$ derivatives. Denote
this basis of operators by $\widehat{\mathcal{O}}_i$. Then one can
re-define the remaining (infinite set of) operators so that their
expansion in PGBs starts at order higher than $n$,
$\mathcal{O}_a\to\mathcal{O}_a-\sum_ic^i_a\widehat{\mathcal{O}}_i$
where the sum runs to $N(n,d)$. Given a desired precision for a
calculation one can determine the order in the loop and momentum
expansions required to achieve that precision. The latter gives us
directly the required number of derivatives $d$ to be retained. The
number $n$ of PGBs to be retained is a bit more complicated. For a
process that involves $k$ PGBs, an operator with $k+2L$ PGBs can
contribute at $L$-loop order. Therefore, for processes with no more
than $k$ PGBs that require $L$-loop precision and up to $d$
powers of momenta, the basis with $N(k+2L,d)$ operators should be used.

While the above algorithm is quite specific, we have not carried out
that program of renormalization. The reason should be clear: the
algorithm requires making  a specific choice of process to study, or
at least a restriction on the number of PGBs in the processes that
will be considered. So, as explained at the top of this section,   we
have opted instead for the full 1-loop renormalization of the model of 
Sec.~\ref{sec:model} assuming all other possible terms consistent with
symmetries (an infinite set) is absent in the bare Lagrangian.

%----------------------------------------------------------------------

\section{Renormalization}
\label{sec:renormalization}

\subsection{Generalities}
\label{subsec:generalities}
The renormalized Lagrangian is
\begin{equation}
\label{eq:su3lag}
{\cal L}={\cal L}_\phi+{\cal L}_\psi+{\cal L}_{\text{Yuk}}
\end{equation}
where
\begin{align}
{\cal L}_\phi&=
\frac{f^2}{8}\bigg[\Tr\left( D_\mu\Sigma^{\dagger} D^\mu\Sigma\right) 
+\sum_{a,i}\zeta_a^{g_i}Z_a^{g_i} \mathcal{O}_a^{g_i}
+\mu^{-\epsilon}\kappa_{\Psi} Z_{\Psi} \mathcal{O}_{\Psi}\bigg],\\
{\cal L}_\psi&=Z_{\chi_L}\bar\chi_{LI}i\slashed{D}\chi^I_L
+Z_{q_R}\bar q_{R}i\slashed{D}q_R\nn\\
&\qquad +\mu^{-\epsilon}\sum_{a=1}^5\kappa_{\chi_a} Z_{\chi_a} Z_{\chi_L}\mathcal{O}_{\chi_a} +\mu^{-\epsilon}\sum_{b=1}^5 \kappa_{q_b} Z_{q_b} Z_{q_R}\mathcal{O}_{q_b},\\
{\cal L}_{\text{Yuk}}&=-f\mu^{\epsilon/2}\lambda_1Z_\lambda
(Z_{\chi_L}Z_{q_R})^{1/2}\bar \chi_{Li}\epsilon^{ij}\epsilon^{xy}
\Sigma_{jx}\Sigma_{3y}q_R+\hc\nonumber\\
&\qquad\qquad-\frac{f}2\mu^{\epsilon/2}\lambda'_1Z_{\lambda'}
(Z_{\chi_L}Z_{q_R})^{1/2}\bar u_{L}\epsilon^{jk}\epsilon^{xy}
\Sigma_{jx}\Sigma_{ky}q_R +\hc\nonumber\\
&\qquad\qquad+f\mu^{-\epsilon}\sum_{a=1}^2\kappa_{v_a}Z_{v_a}\mathcal{O}_{v_a}.
\end{align}
and the wavefunction renormalization of the Goldstone bosons is
implicit in\break $\Sigma=\exp(2iZ^{1/2}\Pi/f)\Sigma_0$.  Note that we
have kept the bare $f$ throughout, and it has dimension $1-\epsilon/2$
in dimensional regularization (with $d = 4-\epsilon$). This ensures
that the coefficients of the power expansion of the kinetic terms do
not run (or rather, they all run the same, just according to the
wavefunction of the field $\Pi$).
Since the spurion $S^{abcde}$ includes the Yukawa coupling constants
it has dimension $\epsilon/2$. Therefore, the bare couplings $\kappa$
have dimension $-\epsilon$. We have ignored the $\lambda_2$ term in
the Yukawa Lagrangian as it plays no role in renormalization.  \\

The calculation will require we fix the $U(1)$ charges of all
fields. For the $\Sigma$ fields these are already determined by the
transformation properties under $G_f$, and the fact that $G_w$ is a
subgroup of $G_f$. Since the hypercharges $Y=Y_1+Y_2$ are fixed and
the interactions are invariant under the gauge transformations, there
is only freedom to choose the $U(1)$ charge of one quark field. We
take $q_R$ to have $Y_2$ charge $y$. Then the rest of the charges are fixed:
\begin{equation}
\label{eq:Ycharges}
\begin{aligned}
Y_1(q_R) &= 2/3-y& Y_2(q_R) &= y \\
Y_1(\chi)&=\sfrac{11}{30}-y & Y_2(\chi_L) &= y-\sfrac15\\
Y_1(u_L) &=\sfrac{13}{15}-y& Y_2(u_L)&=y-\sfrac15
\end{aligned}
\end{equation}

The $\zeta_a$ and $\kappa_a$ terms modify the Lagrangian at tree level
and these modifications should be included in our perturbative
computations. However, we intend to take the bare parameters
$\zeta_a$ and $\kappa_a$  to vanish at the end of the calculation. This
is because we want to study the radiative corrections that generate
these terms, even if absent from the bare Lagrangian. Then, while
$\zeta_a$ terms in the tree level Feynman rules can be neglected, the
counterterms, of the form $\zeta_a(Z_a -1)$, do not vanish as
$\zeta_a\to0$ (similarly for $\kappa_a$  terms). The RGE for these
couplings is derived through standard methods.  We use a generic coupling $\zeta$  for the couplings of $\zeta$, $\kappa$ terms.  Taking a log-derivative with respect to $\mu$ of $\zeta^{\text{bare}}=\mu^{\epsilon D_\zeta}Z \zeta$, where $\epsilon D_\zeta$ is the dimension of the bare coupling $\zeta$, we have
\begin{equation}
	\epsilon D_\zeta Z\zeta +\mu\frac{\partial\zeta}{\partial\mu}\left(Z+\zeta \frac{\partial Z}{\partial\zeta}
 \right)
 +\zeta\left(-\frac12\epsilon g +\beta_g\right) 
\frac{\partial  Z_a}{\partial g} =0.
\end{equation}
Here $g$ stands for the collection of Yukawa and gauge coupling
constants, and there is an implicit sum over these. Since $\mu\frac{\partial\zeta}{\partial\mu}$ has a finite limit as $\epsilon\to0$ and $Z$ can be written as
\[
Z=1+\frac{a^{(1)}}{\zeta}\frac1{\epsilon}+O(\epsilon^{-2}),
\] 
where $a^{(1)}=a^{(1)}(g)$ is only a function of the couplings, we have
\begin{align}
	\mu\frac{\partial\zeta}{\partial\mu} &= -\epsilon D_\zeta \zeta + \beta_\zeta,\nn\\
	\label{eq:betakappaformula}
	\beta_{\zeta} &=  -D_\zeta a^{(1)} + \frac12 g \frac{\partial a^{(1)}}{\partial g}. 
\end{align}
We will determine the normalization factors $Z_a$ in the next subsection.

%----------------------------------------------------------------------

\subsection{Matching Counterterms}
\subsubsection{Scalar 2-Point Functions with Arbitrary Scalar Background}
We first consider the $SU(2)_1$ gauge sector.  The
non-trivial 2-point functions for the scalars are
\begin{align}
  &\includegraphics[width=0.35\textwidth]{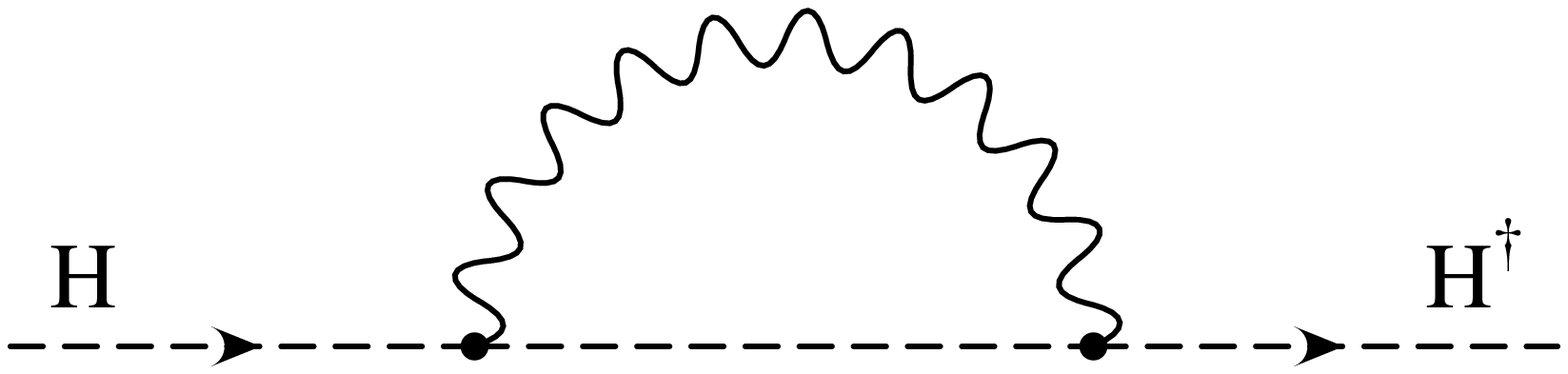}
  \begin{aligned}
    &=  -\frac34 g_1^2 \frac{3}{4}\frac{1}{16\pi^2}\frac{2}{\epsilon}ip^2\delta_{ij}\\
  \end{aligned}
  \displaybreak[0]\\[.6cm]
  &\raisebox{-0.1cm}[0mm][0pt]{\includegraphics[width=0.35\textwidth]{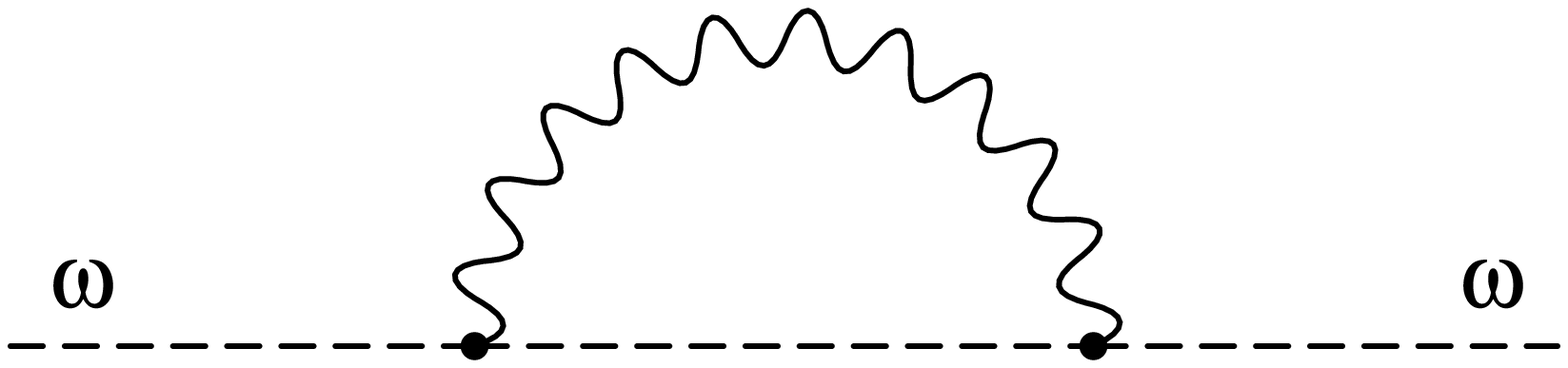}
  }
  \begin{aligned}
    &= -2g_1^2 \frac{3}{4}\frac{1}{16\pi^2}\frac{2}{\epsilon}ip^2
    \delta^{bc}
  \end{aligned}
  \displaybreak[0]\\[.6cm]
  &\raisebox{-.1cm}[5mm][0pt]{\includegraphics[width=0.35\textwidth]{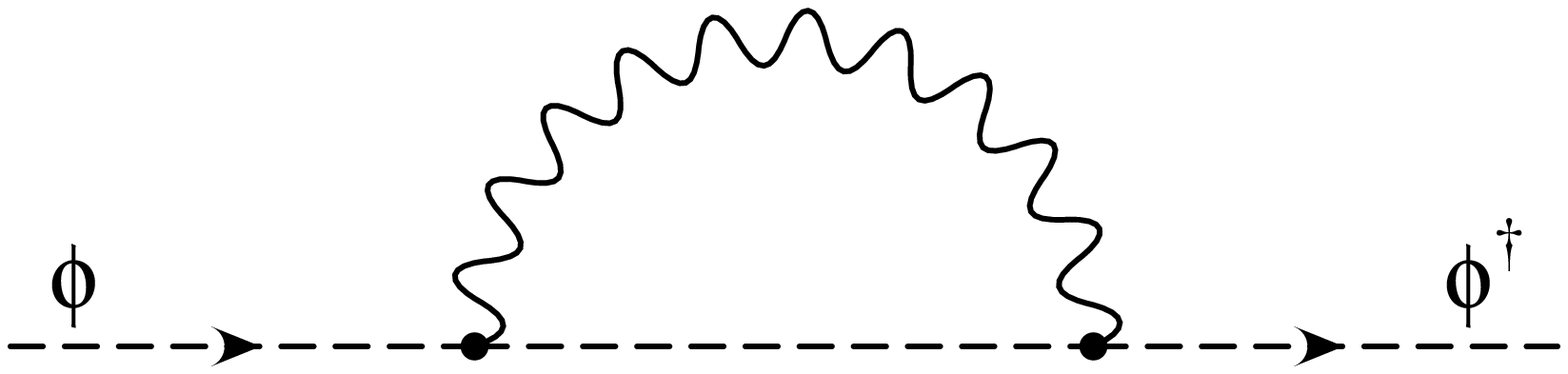}
  }
  \begin{aligned}
    &= -2g_1^2 \frac{3}{4}\frac{1}{16\pi^2}\frac{2}{\epsilon}ip^2
    \delta^{bc}
  \end{aligned}
\end{align}
There is no $\eta$ self-enrgy diagram because it is a singlet under
the gauge group.  For the 3-point functions, we denote by  $p_i$  the
momentum of particle $i$ starting from the left in the clockwise
direction in the following diagrams\\[3mm] 
\begin{align}
  \raisebox{-0.8cm}[0mm][0pt]{\includegraphics[width=0.35\textwidth]{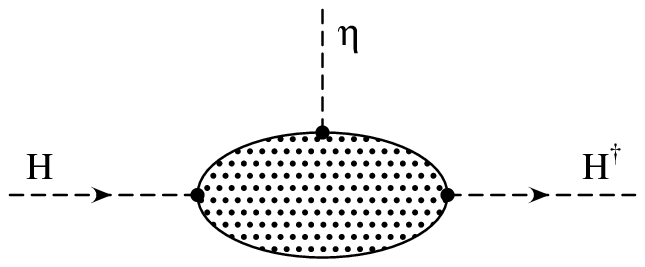}
  } \quad&=\quad\frac{15}{4\sqrt{20}}\frac{g_1^2}{f}
  \frac{3}{4}\frac{1}{16\pi^2}\frac{2}{\epsilon}\left(p_1^2-p_3^2\right)\delta_{ij}
  \displaybreak[0]\\[1cm]
  \raisebox{-1cm}[0mm][0pt]{\includegraphics[width=0.35\textwidth]{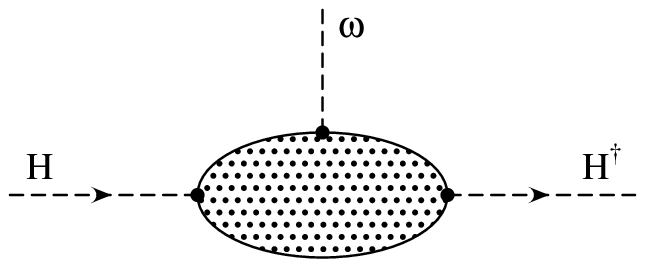}
  } \quad&=\quad\frac54\frac{g_1^2}{f}
  \frac{3}{4}\frac{1}{16\pi^2}\frac{2}{\epsilon}\left(p_1^2-p_3^2\right)\frac{\sigma^a_{ij}}{2}
  \displaybreak[0]\\[1.3cm]
  \raisebox{-0.8cm}[0mm][0pt]{\includegraphics[width=0.35\textwidth]{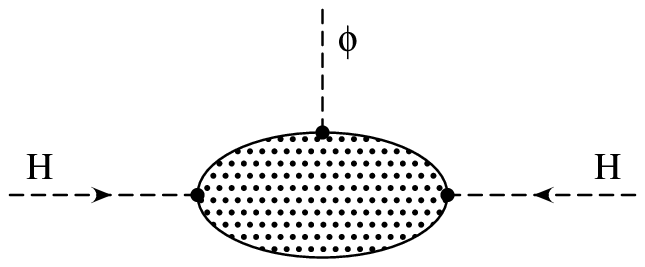} }  \quad&=\quad\frac18\frac{g_1^2}{f} \frac{3}{4}\frac{1}{16\pi^2}\frac{2}{\epsilon}\left(7p_2^2 - 2p_1^2 - 2p_3^2\right)\nn\\
  &\qquad\qquad\times\left(\delta_{ik}\delta_{jl} +
    \delta_{il}\delta_{jk}\right)
  \displaybreak[0]\\[1cm]
  \raisebox{-.6cm}[5mm][0pt]{\includegraphics[width=0.35\textwidth]{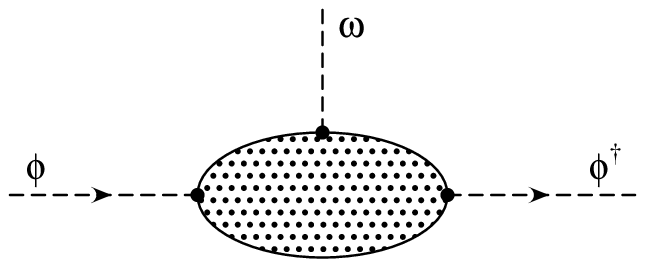}
  } \quad &= \quad
  2\frac{g_1^2}{f}\frac{3}{4}\frac{1}{16\pi^2}\frac{2}{\epsilon}\left(p_1^2-p_3^2\right)
  i\epsilon^{abc}
\end{align}
where we used $\omega = \omega^a\sigma^a/2$ and $\phi =
\sigma^a\sigma^2\phi^a/\sqrt{2}$. We have also found that the 4-point function
with two $\eta$'s and two $\phi$'s vanishes.  Cancelation of divergences
in these diagrams, together with the absence of diagram with two
$\eta$'s and two $\phi$'s requires
\begin{equation}
\begin{split}
  Z &= 1,\\
  Z_1^{g_1} &=  1,\\
  Z_2^{g_1} &= 1+3\frac{1}{\zeta_2^{g_1}}\frac{g_1^2}{16\pi^2}\frac{2}{\epsilon},\\
  Z_3^{g_1} &=
  1+3\frac{1}{\zeta_3^{g_1}}\frac{g_1^2}{16\pi^2}\frac{2}{\epsilon},
\end{split}
\qquad
\begin{split}
  Z_4^{g_1} &=  1,\\
  Z_5^{g_1} &= 1-3\frac{1}{\zeta_5^{g_1}}\frac{g_1^2}{16\pi^2}\frac{2}{\epsilon},\\
  Z_6^{g_1} &= 1+ \frac{3}{20}\frac{1}{\zeta_6^{g_1}}\frac{g_1^2}{16\pi^2}\frac{2}{\epsilon},\\
  Z_7^{g_1} &= 1-\frac32\frac{1}{\zeta_7^{g_1}}\frac{g_1^2}{16\pi^2}\frac{2}{\epsilon},\\
\end{split}
\end{equation}
We can similarly determine $Z_i^{g_2}$ by nothing that
$\mathcal{L}_{g_2}\to \mathcal{L}_{g_1}$ and
$\mathcal{O}_i^{g_2}\to\mathcal{O}_i^{g_1}$ when $\Pi\to-\Pi$.  Thus
we have $Z_i^{g_1} = Z_i^{g_2}$.

We next consider  the $U(1)_1$ sector.  The divergent 2-point and
3-point functions are\\[5mm] 
\begin{align}
  \raisebox{-0.3cm}[0mm][0pt]{\includegraphics[width=0.35\textwidth]{scalar2pt_g_h}
  }\quad &= \quad -\frac14 g^{\prime2}
  \frac{3}{4}\frac{1}{16\pi^2}\frac{2}{\epsilon}ip^2\delta_{ij}
  \displaybreak[0],\\[.6cm]
  \raisebox{-0.3cm}[0mm][0pt]{\includegraphics[width=0.35\textwidth]{scalar2pt_g_phi}
  } \quad &= \quad-g^{\prime2}
  \frac{3}{4}\frac{1}{16\pi^2}\frac{2}{\epsilon}ip^2 \delta^{ab}
  \displaybreak[0],\\[.6cm]
  \raisebox{-0.8cm}[0mm][0pt]{\includegraphics[width=0.35\textwidth]{scalar3pt_g_hhn_c}
  } \quad&=\quad\frac{\sqrt{5}}{8}\frac{g^{\prime2}}{f}
  \frac{3}{4}\frac{1}{16\pi^2}\frac{2}{\epsilon}\left(p_1^2-p_3^2\right)\delta_{ij}
  \displaybreak[0],\\[1cm]
  \raisebox{-1cm}[0mm][0pt]{\includegraphics[width=0.35\textwidth]{scalar3pt_g_hho_c}
  } \quad&=\quad-\frac14\frac{g^{\prime2}}{f}
  \frac{3}{4}\frac{1}{16\pi^2}\frac{2}{\epsilon}\left(p_1^2-p_3^2\right)\frac{\sigma^a_{ij}}{2}
  \displaybreak[0],\\[1.4cm]
  \raisebox{-0.8cm}[0mm][0pt]{\includegraphics[width=0.35\textwidth]{scalar3pt_g_hhp_c}
  } \quad&=\quad-\frac38\frac{g^{\prime2}}{f}
  \frac{3}{4}\frac{1}{16\pi^2}\frac{2}{\epsilon}p_2^2\left(\delta_{ik}\delta_{jl}
    + \delta_{il}\delta_{jk}\right).
\end{align}\\[.2cm]
As in the case of $SU(2)$, the 4-point function with two $\eta$'s and
two $\phi$'s vanishes.  Cancelation of divergences in these diagrams,
together with the absence of a divergence in the diagram with two
$\eta$s and two $\phi$s, requires
\begin{equation}
\begin{split}
  Z_1^{g^\prime_1} &=  1-\frac{1}{40}\frac{1}{\zeta_1^{g^\prime_1}}\frac{g_1^{\prime2}}{16\pi^2}\frac{2}{\epsilon},\\
  Z_2^{g^\prime_1} &= 1+\frac38\frac{1}{\zeta_2^{g^\prime_1}}\frac{g_1^{\prime2}}{16\pi^2}\frac{2}{\epsilon},\\
  Z_3^{g^\prime_1} &= 1+\frac38\frac{1}{\zeta_3^{g^\prime_1}}\frac{g_1^{\prime2}}{16\pi^2}\frac{2}{\epsilon},\\
\end{split}
\qquad
\begin{split}
  Z_4^{g^\prime_1} &=  1,\\
  Z_5^{g^\prime_1} &= 1-\frac38\frac{1}{\zeta_5^{g^\prime_1}}\frac{g_1^{\prime2}}{16\pi^2}\frac{2}{\epsilon},\\
  Z_6^{g^\prime_1} &= 1- \frac{3}{80}\frac{1}{\zeta_6^{g^\prime_1}}\frac{g_1^{\prime2}}{16\pi^2}\frac{2}{\epsilon},\\
  Z_7^{g^\prime_1} &= 1-\frac3{16}\frac{1}{\zeta_7^{g^\prime_1}}\frac{g_1^{\prime2}}{16\pi^2}\frac{2}{\epsilon},\\
\end{split}
\end{equation}
and $Z_i^{g^\prime_1} = Z_i^{g^\prime_2}$. The divergence in the
$H$ 2-point function from Yukawa interaction is
\[\frac{2\lambda_1^2}{16\pi^2}\frac{2}{\epsilon}\]
and in the $\eta$ 2-point function is
\[\frac{8}{5}\frac{\lambda_1^{\prime2}}{16\pi^2}\frac{2}{\epsilon}.\]
Thus we obtain
\begin{equation}
	Z_\Psi = 1-\frac{1}{\kappa_\psi}\frac{1}{16\pi^2}\frac{2}{\epsilon}.
\end{equation}
%-----------------------------------------------------------------------------------
\subsubsection{Fermion 2-Point Functions}
We first consider the $q_R$ 2-point functions with arbitrary scalar
background.  The 1-loop diagrams are
\begin{align}
  \raisebox{-0.8cm}[0mm][0pt]{\includegraphics[width=0.35\textwidth]{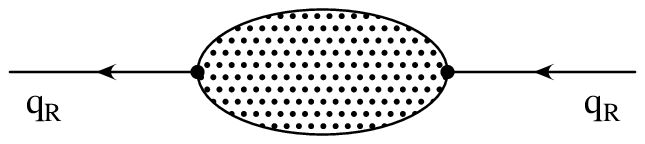}  }\quad &= \quad  \frac{1}{16\pi^2}\frac{2}{\epsilon}\left(2\lambda_1^2+\frac25\lambda_1^{\prime2}\right) i\slashed{p},\\[1.5cm]
  \raisebox{-0.8cm}[0mm][0pt]{\includegraphics[width=0.35\textwidth]{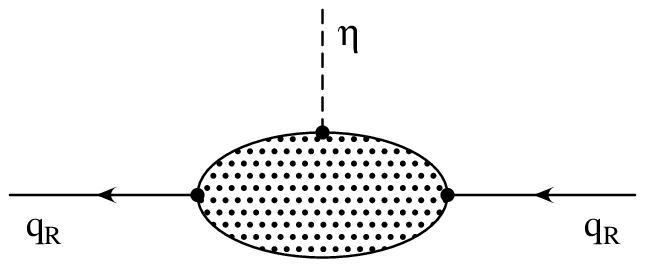}  }\quad &= \quad  \frac{1}{16\pi^2}\frac{2}{\epsilon}\left(-\frac{1}{\sqrt{5}}\lambda_1^2+\frac{4}{5\sqrt{5}}\lambda_1^{\prime2}\right) i\slashed{p}_\eta	\displaybreak[0],\\[1.5cm]
  \raisebox{-0.8cm}[0mm][0pt]{\includegraphics[width=0.35\textwidth]{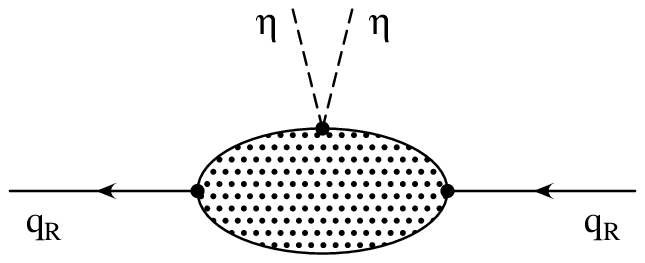}  }\quad &= \quad 0\displaybreak[0],\\[1.5cm]
  \raisebox{-0.8cm}[0mm][0pt]{\includegraphics[width=0.35\textwidth]{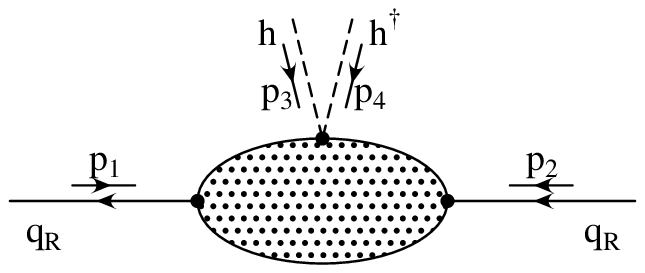}
  } \quad &= \quad \frac{1}{16\pi^2}\frac{2}{\epsilon}\left[
    \frac{3i}{5}(\lambda_1^2-\lambda_1^{\prime2})(\slashed{p}_1-\slashed{p}_2)
    + \frac{i\lambda_1^2}{10}(\slashed{p}_{3}-\slashed{p}_{4})\right].
\end{align}
Matching counterterms yields
\begin{equation}
\begin{split}
	Z_{q_R} &= 1,\\[4mm]
	Z_{q_1} &= 1,
\end{split}
\qquad
\begin{split}
	Z_{q_2} &= 1-8\frac{1}{\kappa_{q_2}}\frac{1}{16\pi^2}\frac{2}{\epsilon},\\
	Z_{q_3} &= 1+\frac85\frac{1}{\kappa_{q_3}}\frac{1}{16\pi^2}\frac{2}{\epsilon},
\end{split}
\qquad
\begin{split}
	Z_{q_4} &= 1,\\[4mm]
	Z_{q_5} &= 1.
\end{split}
\end{equation}
Similarly, for the $\chi_L$ 2-point functions, we find
\begin{equation}
\begin{split}
  Z_{\chi_L} &= 1,\\[4mm]
  Z_{\chi_1} &= 1,
\end{split}
\qquad
\begin{split}
  Z_{\chi_2} &= 1-8\frac{1}{\kappa_{\chi_2}}\frac{1}{16\pi^2}\frac{2}{\epsilon},\\
  Z_{\chi_3} &= 1+\frac85\frac{1}{\kappa_{\chi_3}}\frac{1}{16\pi^2}\frac{2}{\epsilon},
\end{split}
\qquad
\begin{split}
  Z_{\chi_4} &= 1,\\[4mm]
  Z_{\chi_5} &= 1.
\end{split}
\end{equation}

%------------------------------------------------------------------------------------
\subsubsection{Yukawa Vertex Counterterms}
As we mentioned above, gauge interactions do not induce new operators
but the Yukawa interaction do.  Here we distinguish the $SU(2)$ and
$U(1)$ part of the Yukawa in their action:
\begin{equation}
\begin{split}
  Z_{\lambda_1} &= 1-3\frac{1}{16\pi^2}\frac{2}{\epsilon}\left[\left(\frac{11}{30}-y\right)\left(\frac23-y\right)g_1^{\prime2}+\left(y-\frac15\right)yg_2^{\prime2}\right],\\
  Z_{\lambda_1^\prime} &= 1-3\frac{1}{16\pi^2}\frac{2}{\epsilon}\left[\left(\frac{13}{15}-y\right)\left(\frac23-y\right)g_1^{\prime2}+\left(y-\frac15\right)yg_2^{\prime2}\right].\\
\end{split}
\end{equation}
We see that $\lambda_1$ and $\lambda_1^\prime$ are renormalized
differently. The other renomalization factors are:
\begin{equation}
\begin{split}
	Z_{v_1} &= 1+ \frac45\frac{1}{\kappa_{v_1}}\frac{1}{16\pi^2}\frac{2}{\epsilon},\\
	Z_{v_2} &= 1+\frac{11}{5}\frac{1}{\kappa_{v_2}}\frac{1}{16\pi^2}\frac{2}{\epsilon}.
\end{split}
\end{equation}

%----------------------------------------------------------------------
\subsection{$\beta$ Functions}

We have already pointed out that the coupling $\lambda_1$ and
$\lambda_1^\prime$ run differently~\cite{us}.  It is now
straightforward to obtain their beta functions:
\begin{align}
\label{eq:RGElamda1}
\frac{\beta_{\lambda_1}}{\lambda_1} &= 
-\frac3{8\pi^2}\bigg[\left(\sfrac{11}{30}-y\right)\left(\sfrac23-y\right)g_1^{\prime2}
+\left(y-\sfrac15\right)y\;g_2^{\prime2}\bigg]\\
\label{eq:RGElamda2}
\frac{\beta_{\lambda'_1}}{\lambda'_1} &= 
-\frac3{8\pi^2}\bigg[\left(\sfrac{13}{15}-y\right)\left(\sfrac23-y\right)g_1^{\prime2}
+\left(y-\sfrac15\right)y\;g_2^{\prime2}\bigg]
\end{align}
Note, in particular, that
\begin{equation}
\label{eq:RGElmabdas}
\mu\frac{\partial}{\partial\mu}\ln\left(\frac{\lambda_1}{\lambda'_1}\right)=\left(\sfrac23-y\right)\frac{3g_1^{\prime2}}{16\pi^2}
\end{equation}
With $\beta_{g'_1}= (b/16\pi^2)g_1^{\prime3}$ we can write the
solution in terms of the running coupling:
\begin{equation}
  \frac{\lambda_1(\mu)}{\lambda'_1(\mu)}=\frac{\lambda_1(\Lambda)}{\lambda'_1(\Lambda)}
  \left(\frac{g'_1(\mu)}{g'_1(\Lambda)}\right)^{\frac{2-3y}{b}}
\end{equation}
The $\beta$ functions for the couplings $\zeta_a^g$ are determined
using Eq.~\eqref{eq:betakappaformula}.  We find
\begin{equation}
  \begin{aligned}
    \beta_{\zeta_1}^{g^{\PP}_{1(2)}} &= 0,\\
    \beta_{\zeta_2}^{g^{\PP}_{1(2)}} &= 6\frac{g_{1(2)}^2}{16\pi^2},\\
    \beta_{\zeta_3}^{g^{\PP}_{1(2)}} &= 6\frac{g_{1(2)}^2}{16\pi^2},\\
    \beta_{\zeta_4}^{g^{\PP}_{1(2)}} &= 0,\\
    \beta_{\zeta_5}^{g^{\PP}_{1(2)}} &= -6\frac{g_{1(2)}^2}{16\pi^2},\\
    \beta_{\zeta_6}^{g^{\PP}_{1(2)}} &= \frac{3}{10}\frac{g_{1(2)}^2}{16\pi^2},\\
    \beta_{\zeta_7}^{g^{\PP}_{1(2)}} &= -3\frac{g_{1(2)}^2}{16\pi^2},
  \end{aligned}
  \qquad
  \begin{aligned}
    \beta_{\zeta_1}^{g^\prime_{1(2)}} &= - \frac{1}{20}\frac{g_{1(2)}^{\prime2}}{16\pi^2},\\
    \beta_{\zeta_2}^{g^\prime_{1(2)}} &= \frac{3}{4}\frac{g_{1(2)}^{\prime2}}{16\pi^2},\\
    \beta_{\zeta_3}^{g^\prime_{1(2)}} &= \frac{3}{4}\frac{g_{1(2)}^{\prime2}}{16\pi^2},\\
    \beta_{\zeta_4}^{g^\prime_{1(2)}} &= 0,\\
    \beta_{\zeta_5}^{g^\prime_{1(2)}} &= - \frac{3}{4}\frac{g_{1(2)}^{\prime2}}{16\pi^2},\\
    \beta_{\zeta_6}^{g^\prime_{1(2)}} &= - \frac{3}{40}\frac{g_{1(2)}^{\prime2}}{16\pi^2},\\
    \beta_{\zeta_7}^{g^\prime_{1(2)}} &= -
    \frac{3}{8}\frac{g_{1(2)}^{\prime2}}{16\pi^2}.
  \end{aligned}
\end{equation}
For $\kappa$  the couplings are implicit in the
operators so the $\beta$ functions are pure number
\begin{equation}
\begin{aligned}
  \beta_{\kappa_{q_1}} &= 0,\\
  \beta_{\kappa_{q_2}} &= -16\frac{1}{16\pi^2},\\
  \beta_{\kappa_{q_3}} &= \frac{16}{5}\frac{1}{16\pi^2},\\
  \beta_{\kappa_{q_4}} &= 0,\\
  \beta_{\kappa_{q_5}} &= 0,\\
\end{aligned}
\qquad
\begin{aligned}
  \beta_{\kappa_{\chi_1}} &= 0,\\
  \beta_{\kappa_{\chi_2}} &= -16\frac{1}{16\pi^2},\\
  \beta_{\kappa_{\chi_3}} &= +\frac{16}{5}\frac{1}{16\pi^2},\\
  \beta_{\kappa_{\chi_4}} &= 0,\\
  \beta_{\kappa_{\chi_5}} &= 0,\\
\end{aligned}
\end{equation}
and
\begin{equation}
  \beta_\psi = -2\frac{1}{16\pi^2},\quad
  \beta_{\kappa_{v_1}} = \frac{8}{5}\frac{1}{16\pi^2},\quad
  \beta_{\kappa_{v_2}} = \frac{22}{5}\frac{1}{16\pi^2}.
\end{equation}

%----------------------------------------------------------------------

\section{Conclusions}
\label{sec:conc}
We have studied the one loop renormalization of the Littlest Higgs
Model. Phenomenologically this model has fallen  somewhat out of favor
because of its difficulties simultaneously accommodating electroweak
precision constraints and solving the little hierarchy problem
\cite{Han:2003wu}.  However, its structure is prototypical of many models,
like Littlest Higgs models with reduced gauge symmetry
\cite{Perelstein:2003wd}, or with custodial \cite{Chang:2003un} or
T-parity \cite{Cheng:2003ju} symmetries. Therefore, the methods
introduced here should be largely the same as those needed 
for one loop renormalization
of any model in this class.

We have displayed explicit counterterms and their $Z$
factors in dimensional regularization, in Landau gauge. These results
are only of interest to understand the procedure, so they have been
included here more for clarity of presentation. However, the beta
functions of the couplings of all the terms in the Lagrangian are
independent of gauge and scheme choice. They, together with the
methods introduced, constitute the main result of this work and are
displayed explicitly in Sec.~\ref{sec:renormalization}.

One important result is that the coupling constants associated with
the Yukawa coupling of the top quark run differently; see
Eq.~\eqref{eq:RGElmabdas}. As observed in Ref.~\cite{us} in the
absence of fine tuning, the collective symmetry mechanism fails for
Yukawa couplings in the Littlest Higgs model and its relatives. One
can similarly conclude that the terms that were required as
counterterms, all allowed by the symmetries and being of leading order
in the derivative expansion, should have been included in the model
from the start.

%----------------------------------------------------------------------

\vspace{1cm}
\begin{center}
  {{\bf{Acknowledgments}}}
\end{center}
Work supported in part by the US Department of Energy under contract
DE-FG03-97ER40546.

\end{document}